# Differentiating Contributions of Electrons and Phonons to the Thermoreflectance Spectra of Gold

Kexin Liu[1], Xinping Shi[1], Frank Angeles[1], Ramya Mohan[2],
Jon Gorchon[3], Sinisa Coh[1,2,a], Richard B. Wilson[1,2,b]

1) Mechanical Engineering, University of California, Riverside, CA, 92521, USA
2) Materials Science and Engineering, University of California, Riverside, CA, 92521, USA
3) Université de Lorraine, CNRS, IJL, F-54000 Nancy, France
a) sinisa.coh@gmail.com
b) rwilson@ucr.edu

*Abstract*

To better understand the many effects of temperature on the optical properties of metals, we experimentally and theoretically quantify the electron vs. phonon contributions to the thermoreflectance spectra of gold. We perform a series of pump/probe measurements on nanoscale Pt/Au bilayers at wavelengths between 400 and 1000 nm. At all wavelengths, we find that changes in phonon temperature, not electron temperature, are the primary contributor to the thermoreflectance of Au. The thermoreflectance is most sensitive to the electron temperature at a wavelength of ~480 nm due to interband transitions between d-states and the Fermi-level. At 480 nm, the electron temperature is responsible for ~20% of the total thermoreflectance. In the near infrared, the electron temperature is responsible for < 2% of the total thermoreflectance. We also compute the thermoreflectance spectra of Au from first principles. Our calculations further confirm that phonon temperature dominates thermoreflectance of Au. Most of Au's thermoreflectance is due to the effect of the phonon population on electron lifetime.

## I. Introduction

Despite a half-century of study [1-15], the relative contributions of electrons vs. phonons to the temperature dependence of a metal's optical properties are not clear. A relatively common explanation for thermo-optic spectra of metals is that the electron temperature affects interband optical transitions [16-19]. An interband transition threshold is the energy difference between band extrema and the Fermi-level. Absorption probabilities are higher for photon energies near the transition threshold. Therefore, a change in the electronic occupancy near the Fermi-level can increases/decrease absorption for photons $k_B T$ above/below the interband transition threshold. This electron-temperature-based explanation for the thermoreflectance spectra is often invoked to explain the thermoreflectance spectra of simple metals Cu, Ag, and Au [19,20]. These metals have a maxima in their thermo-optic spectra near their interband transition thresholds. Other ways for the electron temperature to affect optical properties include altering electron-electron scattering rates [21], and shifting the Fermi-level [1].

An alternative explanation for thermo-optic spectra of metals is thermal expansion [5]. Like changes in the electron occupancy, thermal expansion can also change the optical transition probabilities between band extrema and the Fermi-level [5]. Since thermal expansion occurs as a result of increasing phonon populations, this explanation credits thermo-optic spectra to phonons. Additionally, electron-phonon interactions can affect optical properties [9,14]. Electron-phonon scattering rates are determined by the phonon temperature.

Experimentally evaluating the relative importance of electrons vs. phonons to the thermo-optic spectra of metals is challenging. While many experimental studies report the thermoreflectance spectra of metals [1-15], these experiments were carried out at conditions where electrons and phonons were in equilibrium. To differentiate electron and phonon contributions to thermo-optic properties, the metal needs to be driven into a nonequilibrium state where electron and phonon temperatures differ, $T_e \neq T_p$. Additionally, the nonequilibrium state needs to be such that the electron and phonon temperatures are definable. In other words, the electron distribution needs to be a Fermi-Dirac distribution, and the phonon distribution needs to be a Bose-Einstein distribution.

A straightforward way to cause nonequilibrium between electrons and phonons is photoexcitation [22]. However, photoexcitation drives the metal into a nonequilibrium state where the electrons and/or phonons have nonthermal distributions [18,23-26]. The distributions remain nonthermal until enough electron-electron, electron-phonon, and phonon-phonon scattering events occur to maximize entropy. Depending on the scattering rates, it takes between 0.1 and 5 picoseconds for a nonthermal distribution to evolve into a thermal one [17,24,27]. This timescale is typically comparable to how long it takes for electrons and phonons to thermalize with each other [24]. In short, after photoexcitation, the electron and phonon temperatures may not be well defined until $T_e \approx T_p$. Several recent studies report that the nonthermal character of photoexcited electron distributions affects the optical response of metals in pump/probe experiments [17,18].

To overcome the challenge of nonthermal effects, we perform pump/probe measurements on Au/Pt bilayers. After photoexcitation of the Pt layer, the electron and phonon temperatures in the Au are both definable, and unequal, on timescales between 2 and 100 ps after photoexcitation [28]. Unlike a single Au layer, where electrons and phonons take only a few ps to thermalize after photoexcitation [17], electrons and phonons in a Au/Pt bilayer take as long as 100 ps to equilibrate [28,29]. We show the temperature dynamics following photoexcitation for a Au/Pt bilayer in Fig. 1. In a single Au film, the electron-phonon thermalization time is short because Au electrons have a small heat-capacity. Alternatively, in a Au/Pt bilayer, the Au electrons, Pt electrons, and Pt phonons have strong enough thermal coupling to effectively form a single thermal reservoir [28,29]. Collectively, this reservoir has a heat-capacity two-orders of magnitude larger than Au electrons in isolation, and therefore takes ~100 times longer to thermalize with Au phonons. Another advantage of Au/Pt bilayers is that strong electron-electron interactions in Pt reduce the importance of nonthermal effects to the dynamics at all timescales [24,30].

The main results of our study are wavelength-dependent pump/probe measurements of Au/Pt bilayers. Our wavelength-dependent experiments reveal that the thermoreflectance of Au is primarily driven by changes in phonon temperature. The thermoreflectance of Au is most sensitive to electrons at 480 nm, where $dR/dT_e$ makes up ~20% of the total thermoreflectance. Alternatively, in the near-infrared, the electron temperature is responsible for < 2% of the total thermoreflectance.

We also computed the thermoreflectance spectra of Au from first principles. Accurately calculating the thermoreflectance is challenging because of the multitude of ways that temperature affects optical properties. Nevertheless, in agreement with our experimental data, our first-principles calculations find that the dominant contribution to thermoreflectance spectra is through phonons. The large thermoreflectance of Au near the interband transition threshold is dominated by the temperature dependence of the electron lifetime due to electron-phonon interactions. Increases in the electron-phonon scattering rate decrease/increase absorption at wavelengths shorter/longer than the interband transition threshold wavelength of 510 nm.

**II. Experimental Methods.**

We deposited a 60 nm Au thin film, and two Au/Pt bilayers on sapphire substrates for pump/probe experiments using a direct current magnetron sputter deposition system. The two bilayer samples are the focus of our study. The geometry of the two bilayer samples are (71nm Au)/(17nm Pt)/sapphire, and (64 nm Au)/(19nm Pt)/sapphire. Electron backscattering analyses show that the films are highly (111) textured. Further information on the sample preparation, and our measurements of film thickness are provided in [31].

We perform front/back time-domain thermoreflectance measurements on the Pt/Au bilayers. By "front/back" we mean the probe beam measures the Au reflectance at the front of the sample (Au/air interface), and the pump photoexcites the Pt layer through the back of the sample (Pt/sapphire interface). The pulse duration of the pump and probe beams is ~100-200 fs. The pump beam is electro-optically modulated with a 50% duty cycle at 10.7 MHz. The intensity of the reflected probe pulse is monitored with an amplified silicon photodiode detector. The photodetector is connected to a radio frequency lock-in the detects 10.7 MHz signal. Both pump and probe beams have a $1/e^2$ radius of ~7 µm. We conduct experiments where the pump and probe beams have the same wavelength, tuned between 690 and 1000 nm. To extend the range of our study, we also conduct experiments with a frequency-doubled probe beam with wavelengths between 400 and 525 nm. Further details of our experimental setup are reported in Ref. [32].

We analyze our pump/probe data with a two-temperature thermal model. The predictions of our thermal model for the (71nm Au)/(17nm Pt) sample are summarized in Fig. 1. The model consists of two coupled heat equations for each layer in our sample. The heat equations describe how the electron and phonon temperatures evolve in space and time after photoexcitation with the pump.

$$C_e \frac{\partial T_e}{\partial t} = \Lambda_e \frac{\partial^2 T_e}{\partial z^2} + g_{ep}\left(T_p - T_e\right) + S(z,t). \qquad (1)$$

$$C_p \frac{\partial T_e}{\partial t} = \Lambda_p \frac{\partial^2 T_p}{\partial z^2} + g_{ep}(T_e - T_p). \tag{2}$$

Here $T_e$ and $T_p$ are the electron and phonon temperatures, $C_e$ and $C_p$ are the electron and phonon heat-capacities, $g_{ep}$ is the electron-phonon energy transfer coefficient. The model takes into consideration where the heat is deposited via $S(z,t)$, which describes the absorption of optical energy by electrons (Fig 1).

The input parameters of the thermal model are the thermal properties and optical constants of the sample. The outputs of the thermal model are the electron and phonon temperatures as a function of depth and time. All thermal model parameters we use to calculate the evolution of heat in the bilayer are fixed by independent measurements, or literature values. To fix electronic thermal conductivities in our model, we sputtered separate 60 and 200 nm thick Pt and Au single-layer films on sapphire. We then used the four-point probe method to measure the electrical resistivity of these films. Then, using the Wiedemann Franz law, we estimate $\Lambda_e \approx 190$ W m$^{-1}$ K$^{-1}$ for Au and $\Lambda_e \approx 40$ W m$^{-1}$ K$^{-1}$ for Pt. We assume that the electronic heat capacity scales linearly with electron temperature, $C_e = \gamma T_e$. For Au, based on Ref. [29], we use $\gamma_{Au} \approx 70 \cdot 10^{-6}$ J m$^{-3}$ K$^{-2}$, $g_{ep} \approx 2 \cdot 10^{16}$ W m$^{-3}$ K$^{-1}$, $C_p \approx 2.5$ MJ m$^{-3}$ K$^{-1}$, and $\Lambda_p \approx 3$ W m$^{-1}$ K$^{-1}$. For Pt, based on Ref. [30], we use $\gamma_{Pt} \approx 4 \cdot 10^{-4}$ J m$^{-3}$ K$^{-2}$, $g_{ep} \approx 6 \cdot 10^{17}$ W m$^{-3}$ K$^{-1}$, $C_p \approx 2.7$ MJ m$^{-3}$ K$^{-1}$, and $\Lambda_p \approx 7$ W m$^{-1}$ K$^{-1}$. The shaded region in Fig. 2 indicates what range of electron and phonon temperatures are possible given uncertainties in the model's input parameters.

We assume the time-dependence of $S(z,t)$ tracks the intensity vs. time of our laser pulse, which we previously measured and reported as a function of wavelength in Ref. [32]. Zero-delay time is defined as the time when the pump beam intensity is a maximum. This means that the pump pulse deposits half of its energy before zero delay-time. To determine the depth-dependence of $S(z,t)$, we calculate the absorption profile using a multilayer reflectivity calculation similar to the one described in Ref. [29]. We use literature values for the indices of refraction of similarly prepared Au and Pt thin-films [11]. The depth dependence of the absorption of the pump is shown in the top panel of Fig. 1 for a pump wavelength of 783 nm. In our experiments, as we change the laser wavelength, the depth dependence of $S(z,t)$ is slightly altered due to the wavelength dependence of the index of refraction of Au and Pt. However, at all wavelengths that we conducted experiments, at least 97% of the pump pulse's energy is absorbed directly by the Pt layer.

Solving Eq. (1) and (2) requires boundary conditions. We couple phonon heat-equations for Pt, Au, and sapphire by assuming the heat-current carried by phonons across the interface is $J_p = G_{int} \Delta T_p$, where $\Delta T_p$ is the phonon temperature difference across the interface, and $G_{int}$ is the interfacial thermal conductance for either the Au/Pt and Pt/Sapphire interface, respectively. We take $G_{int}$ values from Ref. [29]. For boundary conditions on the electron heat-equations, we assume an adiabatic boundary condition at the Pt/Sapphire and Au/air interfaces. For electrons at the Au/Pt interface, the model assumes $J_e = G_{e-e} \Delta T_e$. Data for the electronic thermal conductance of Au/Pt interfaces is limited [29], but measurements of specific electrical resistance combined with the Wiedemann Franz law [33] predict $G_{e-e}$ of 10 and 30 GW m$^{-2}$ K$^{-1}$ for Pt/Cu and Pd/Au interfaces [34,35]. Our model predicts no detectable change in the temperature evolution of Au electrons and phonons for values of $G_{e-e}$ ranging between 5 GW m$^{-2}$ K$^{-1}$ and infinity. Therefore, for simplicity, we set $G_{e-e} = \infty$, i.e. we assume the electron temperature is continuous across the Au/Pt interface.

Our model predicts that heat is transported into the Au phonons in two ways. First, heat is exchanged between Au electrons and phonons due to the $g_{ep}$ term in Eqs. (1-2). Second, heat can diffuse from the Pt phonons into the Au phonons, due to a conductive boundary condition on the heat-current at the Au/Pt interface (see [31] for more details). To qualitatively gauge the relative importance of these two heat-transfer mechanisms, we estimate the thermal conductance for these two processes. The effective conductance per unit area between Au electrons and Au phonons is $g_{ep}^{Au} d_{Au} \approx 1$ GW m$^{-2}$ K$^{-1}$, where $d_{Au}$ is the length-scale over which Au electrons and phonons are different temperatures. For a thick metal layer, this nonequilibrium length-scale is $\approx \sqrt{\Lambda_e / g_{ep}}$, which is ~ 100 nm for Au [29]. In our samples, the nonequilibrium length-scale $d_{Au}$ is limited by the ~70 nm thickness of the Au film. In contrast to the large electron-phonon conductance, a typical value for the phonon-phonon conductance at an interface between two materials is only between 0.1 to 0.3 GW m$^{-2}$ K$^{-1}$ [36,37]. Therefore, we conclude electron-phonon energy exchange is the primary mechanism by which Au phonons are heated.

To compare to our experimental data, we parameterize the change in reflectance as

$$\Delta R(t) = C_{TR}(\lambda)\left[a(\lambda)\Delta T_e(t) + b(\lambda)\Delta T_p(t)\right] . \qquad (3)$$

Here, $t$ is the time-delay between the pump and probe pulses, $\Delta T_e(t)$ is the transient electron temperature, $\Delta T_p(t)$ is the transient phonon temperature, and $C_{TR}(\lambda)$ describes the dependence of the thermoreflectance spectra on wavelength $\lambda$. We define $C_{TR}$ as the magnitude of the equilibrium thermoreflectance coefficient, $C_{TR}(\lambda) = |dR/dT|$ when $\Delta T_e = \Delta T_p$. The functions $a(\lambda)$ and $b(\lambda)$ define the sensitivity of the thermoreflectance to electrons vs. phonons. The relationship between these sensitivity functions and the partial derivatives of the reflectance is $\partial R/\partial T_e = a(\lambda)C_{TR}(\lambda)$ and $\partial R/\partial T_{ph} = b(\lambda)C_{TR}(\lambda)$. The value of $a(\lambda) + b(\lambda)$ must equal 1 or -1, depending on whether the thermoreflectance is positive or negative at that $\lambda$. The temperatures $\Delta T_e(t)$ and $\Delta T_p(t)$ are a weighted average of the electron and phonon temperature profile as a function of depth [29]. The weighted average is calculated using a multilayer reflectivity calculation [38] that we describe in [31].

We do not consider the temperature dependence of $C_{TR}(\lambda)$. Prior studies suggest the thermoreflectance of Au is nearly constant across temperatures between 300 and 500 K [15]. The temperature rise of the Au electrons in our experiment is less than 200 K on timescales less than 3 ps, and less than 20 K on the 3 to 200 ps timescales we fit our data across.

### III. Theoretical Methods.

We use the first principles calculated electron band structure energies $E_{n\mathbf{k}}$ and orbitals $\psi_{n\mathbf{k}}$ within the perturbative approach to evaluate the optical conductivity of Au [39,40],

$$\sigma_{\alpha\beta}(\omega) = \frac{i e^2 \hbar}{(2\pi)^3} \lim_{q \to 0} \int d\mathbf{k} \sum_{n,m} \frac{f_{m\mathbf{k}+\mathbf{q}} - f_{n\mathbf{k}}}{E_{m\mathbf{k}+\mathbf{q}} - E_{n\mathbf{k}}} \frac{\langle\psi_{n\mathbf{k}}|v_\alpha|\psi_{m\mathbf{k}+\mathbf{q}}\rangle \langle\psi_{m\mathbf{k}+\mathbf{q}}|v_\beta|\psi_{n\mathbf{k}}\rangle}{E_{m\mathbf{k}+\mathbf{q}} - E_{n\mathbf{k}} - \hbar\omega - i\eta_{mn\mathbf{k}}/2}, \qquad (4)$$

The conductivity $\sigma_{\alpha\beta}$ describes the current in direction $\alpha$ in response to an electric field pointing in direction $\beta$. Eq. (4) is a summation over possible electronic transitions between states in band $m$ at wavevector $\mathbf{k}+\mathbf{q}$ to states in band $n$ and wavevector $\mathbf{k}$. The limit as $\mathbf{q} \to 0$ indicates that we include both intraband and interband contributions to the optical conductivity. The Fermi-Dirac distribution occupation factor is denoted as $f$ while the velocity operator is denoted by $v$. $\eta_{mn\mathbf{k}}$

describes the effect of electronic scattering rates on transitions due to electron-electron scattering. We treat the electron-phonon scattering via the special displacement method, described in [41].

Electron and phonon temperature appear at several places in the expression above for the optical conductivity. For example, the electron temperature $T_e$ appears in the occupation factors, $f_{nk} \to f_{nk}(T_e)$. However, the electron temperature also affects optical properties by altering electron-electron scattering rates, $\eta_{mnk} \to \eta_{mnk}(T_e)$. Changes in phonon population alter $\psi_{nk}$ and $E_{nk}$ because electron-phonon interactions shift and warp energy bands (see Fig. 3). The phonon population also changes optical properties by modifying electron-phonon scattering rates. Finally, the phonon temperature $T_p$ also affects Eq. (4) through the effect of volume expansion $V(T_p)$ on $\psi_{nk}$ and $E_{nk}$.

We treat the effect of phonon temperature on optical properties with the special displacement method approach described in [41] and [42]. In this approach, Au atoms in a supercell are displaced by vector $\xi$ away from their equilibrium locations based on the phonon temperature $T_p$. As described in [41], vector $\xi$ is computed as a sum over phonon eigenvectors, weighted by the Bose-Einstein occupation factor. Once we obtain the band-structure for a system with displaced atoms we recompute its optical conductivity. The special displacement method approach includes both Debye-Waller and Fan-Migdal electron-phonon terms, and accurately describes the effects of electron-phonon interactions on both the real and imaginary part of the interband and intraband optical conductivity. The inclusion of both Debye-Waller and Fan-Migdal terms is an advantage over conventional treatments of electron-phonon interactions [14,43]. Similarly, we treat the effect of phonon-driven thermal expansion by computing the band structure and optical conductivity with different volume unit-cells, $\psi_{nk} \to \psi_{nk}[V(T_p)]$ and $E_{nk} \to E_{nk}[V(T_p)]$.

We model the temperature dependence of the electron-electron interaction contribution to carrier scattering rates using the Fermi liquid approach [44],

$$\eta_{mnk} = 0.0082 \text{ eV}^{-1} \left[ \frac{1}{2}(E_{mk} - E_F)^2 + \frac{1}{2}(E_{nk} - E_F)^2 + (\pi k_B T_e)^2 \right], \quad (5)$$

where we average the contribution of electron and hole state. We arrived at a prefactor of $0.0082 \text{ eV}^{-1}$ by fitting to the first-principles GW calculations [45-47]. This prefactor is also consistent with two-photon photoemission data [48-50]. Importantly, Eq. (5) does not explicitly include the effects of electron-phonon interactions. Those effects are already included in our previously mentioned special displacement method approach.

Using the framework described above, we compute the thermoreflectance due to electron ($T_e$) or phonon ($T_p$) temperature with a finite-difference approach. Electron wavefunctions and energies are computed within the PBEsol+U approach as implemented in the quantum espresso package [51]. We use U=2.7 eV following Ref. [43]. We sample the indirect absorption on a supercell containing 64 gold atoms. We sample the charge density on a 24x24x24 mesh of k-points in the equivalent one-atom unit cell. The interband contributions to the optical conductivity are well converged with a 200x200x200 mesh of k-points in the equivalent one-atom unit cell. The intraband contributions converge at 40x40x40 mesh.

**IV. Results and Discussion**.

The change in reflectance of the Au layer after pump heating of the adjacent Pt layer is shown in Fig. 2 for probe wavelengths of 480, 695, and 960 nm. For wavelengths in the near-infrared, $\Delta R$ slowly increases for ~100 ps after excitation. Alternatively, for wavelengths near the interband transition threshold of Au, e.g. 480 nm, $\Delta R$ depends only weakly on time after the first few picoseconds. We credit these differences in thermoreflectance signals at timescales between 2 and 100 ps to differences in sensitivity to the electron vs. phonon temperatures.

We fit the data with Eq. (3) by treating the electron and phonon sensitivity parameters $a$ and $b$ as fit parameters at each wavelength, see Fig. 2. Figure 4 summarizes the best-fit values for $a(\lambda)$ at wavelengths between 400 and 1000 nm. The markers are the average value of $a(\lambda)$ we deduce by fitting data from both Pt/Au bilayer samples. The error bars account for the uncertainties in $\Delta T_p(t)$ and $\Delta T_e(t)$ due to uncertainties in the thermal model input parameters. Most of the uncertainty in $a(\lambda)$ arises from a 10% uncertainty in the thickness of the Pt film.

To test our concerns that non-thermal effects [17,18,24] prevent us from differentiating the effect of $T_e$ vs. $T_p$ on thermoreflectance in a single Au layer, we also performed measurements on a 60 nm Au film with no Pt film. We used the values for $a(\lambda)$ shown in Fig. 4 to make two-temperature model predictions for $\Delta R(t)$ of the Au film. At some wavelengths, predictions for $\Delta R(t < 2\text{ ps})$ disagree with the data in both magnitude and sign, see [31].

The values for $a(\lambda)$ and $b(\lambda)$ we derive from our experiments are intrinsic properties to Au, and do not depend on the details of our sample geometry. The reflectance of the probe beam depends primarily on the optical properties of the Au layer because the Au thickness is much greater than the optical penetration depth. The optical penetration depth of Au is 16 and 12 nm for wavelengths of 400 and 1000 nm, respectively. However, there are uncertainties in the derived values of $a(\lambda)$ and $b(\lambda)$ due to the accuracy of thermal model prdictions for the temperature evolution of the Au on timescales from 3 to 250 ps. To test the robustness of our thermal model predictions, we performed an additional experiment at 783 nm on a 73nm Au/17nm-Fe sample. Like Pt, Fe has a much stronger electron-phonon energy transfer coefficient then Au. A best fit value to our experimental data with our thermal model predictions for the Au/Fe yields $b(783 \text{ nm}) \approx 0.99$, see [31]. This value is in fair agreement with the value of $b(783 \text{ nm}) \approx -0.98 \pm 0.007$ we derived from the experiments on the two Au/Pt samples, and supports our conclusion that phonons dominate the thermoreflectance spectra.

The results of our first-principles calculated thermoreflectance of Au are reported in Figs. 4 and 5. Overall we find good quantitative and qualitative agreement with the experiment, as can be seen from Figs. 4(b) and (c). Our theory predicts that the total thermoreflectance of gold has a peak value of $-2.5 \times 10^{-4}$ K$^{-1}$, in excellent agreement with experimental studies that report a peak thermoreflectance for Au between -2 and $-3 \times 10^{-4}$ K$^{-1}$ [11,15,52]. We experimentally observe a maximum thermoreflectance at ~540 nm, while theory predicts a maximum near 520 nm. The theoretical predictions deviate from the experimental data at wavelengths below 520 nm, which is the interband transition threshold energy. There are two likely reasons for the discrepancy between experiment and theory at higher energies. First, the density functional theory is a ground-state theory. A fully rigorous description of optical properties requires a first-principles theory for the excited state, such as the GW-BSE approach. We do not use GW-BSE because its computational cost would make parts of our calculation infeasible, e.g. the special displacement method supercell approach for $T_p$ effects. A second reason for discrepancy at high energy is the different elastic boundary condition in the model vs. experiment. In our model, the bulk gold is free to expand in all directions. In the experiment, the gold film is deposited on a substrate that will hinder expansion in the directions parallel to the substrate.

Theory and experiment are also in agreement on how much the electron temperature affects the reflectance of Au, see Fig. 4b. Theory predicts a maximum value of $|\partial R / \partial T_e| \approx 0.5 \cdot 10^{-4} K^{-1}$ for energies above and below the interband transition threshold. To deduce $|\partial R / \partial T_e|$ from our experimental measurements, we multiply $a(\lambda)$ reported in Fig. 4a with $C_{TR}(\lambda)$ reported in Ref. [11]. We arrive at a maximum value of $|\partial R / \partial T_e| \approx 0.3 \cdot 10^{-4} K^{-1}$. Like our theory predictions, the experimental maximum occurs above and below the interband transition threshold energy.

At longer wavelengths, above 600 nm, the contribution of electrons to the total thermoreflectance is even less important. The total thermoreflectance in this energy range is around $-0.3 \cdot 10^{-4} K^{-1}$ in both theory and experiment. The electronic contribution is $\approx -5 \cdot 10^{-7} K^{-1}$ in experiment and between $-10^{-6} K^{-1}$ and $-10^{-7} K^{-1}$ in our calculation. Therefore, we conclude that at all energies thermoreflectance of Au is dominated by the phonon temperature.

The primary way thermally displaced atoms $\xi(T_p)$ effect the dielectric function is through modifications in the electron-phonon lifetime. To confirm this, we did a new calculation with two differences from the approach described above. First, we removed the effect of thermally displaced atoms from our calculation. Second, we added an electron-phonon scattering term to electron lifetime in Eq. (5) that is proportional to $T_p$. We set the proportionality constant based on the temperature-dependence of the electrical conductivity. We found this simple but crude method for including the effects of electron-phonon scattering qualitatively reproduces the thermoreflectance predicted by the special displacement method approach. Our findings in Fig. 5 correct older band-structure studies that concluded thermoreflectance spectra were primarily due to thermal expansion [5,9].

## V. Conclusion.

We have experimentally determined that phonons dominate the thermoreflectance spectra of Au. By experimentally quantifying the sensitivity of the thermoreflectance spectra to electron vs. phonon temperatures, it is now possible to use time-domain thermoreflectance experiments to differentiate between ultrafast electron vs. phonon dynamics in metal films. Previously, such differentiation has only been possible using time-resolved X-ray diffraction experiments [53]. Our first principles calculations show that phonons dominate the thermoreflectance spectra by changing the electron lifetime due to electron-phonon interactions. While our study focuses only on Au, we expect our conclusions will be applicable to other metals for several reasons. Many metals, e.g. Al, Ta, Cu, and Ag, have sharp resonance-like features in the thermoreflectance spectra near interband transition threshold energies. These similarities in thermoreflectance spectra suggest similar origins. Furthermore, most metals have significantly stronger electron-phonon coupling than the heavy-metal Au, since the strength of electron-phonon interactions depends on atomic mass. Stronger electron-phonon interactions will increase the dependence of the dielectric function on phonon temperature.

Our findings are important for understanding and interpreting phenomena in many fields, including nanophotonics and plasmonics [54-56], ultrafast electron dynamics [16,17,57,58], ultrafast magnetism [59-63], and nanoscale heat-transfer [64-66]. Scientists in these fields use optical pump/probe experiments to study charge and energy transport in plasmonic and nanophotonic devices [18,54-56], or nanoscale metal multilayers [28,30,59-63,67-69]. Important length-scales in pump/probe measurements of metal structures and devices are less than 100 nm. Important time-scales are less than 100 ps [28]. At these short length- and time-scales, electrons and phonons in the metal can be out of equilibrium [24,70,71]. Knowledge of how and why a metal's optical properties depend on the phonon vs. electron temperature will aid the interpretation of pump/probe measurements of nonequilibrium phenomena [17,54,65,68,70-72].

**Acknowledgement**

We thank Ivo Souza for his assistance with Wannier90 code. The work by K. L. X. S., R.M., and R.W. was primarily supported by the U.S. Army Research Laboratory and the U.S. Army Research Office under contract/grant number W911NF-18-1-0364. K. L. and R.W. also acknowledge support by NSF (CBET – 1847632). S.C. was supported by NSF (DMR-1848074). We also thank the French PIA project "Lorraine Université d'Excellence" reference ANR-15-IDEX-04-LUE. Work by J.G. was supported as part of project PLUS by the metropole Grand Nancy and the "FEDER-FSE Lorraine et Massif Vosges 2014-2020", a European Union Program.

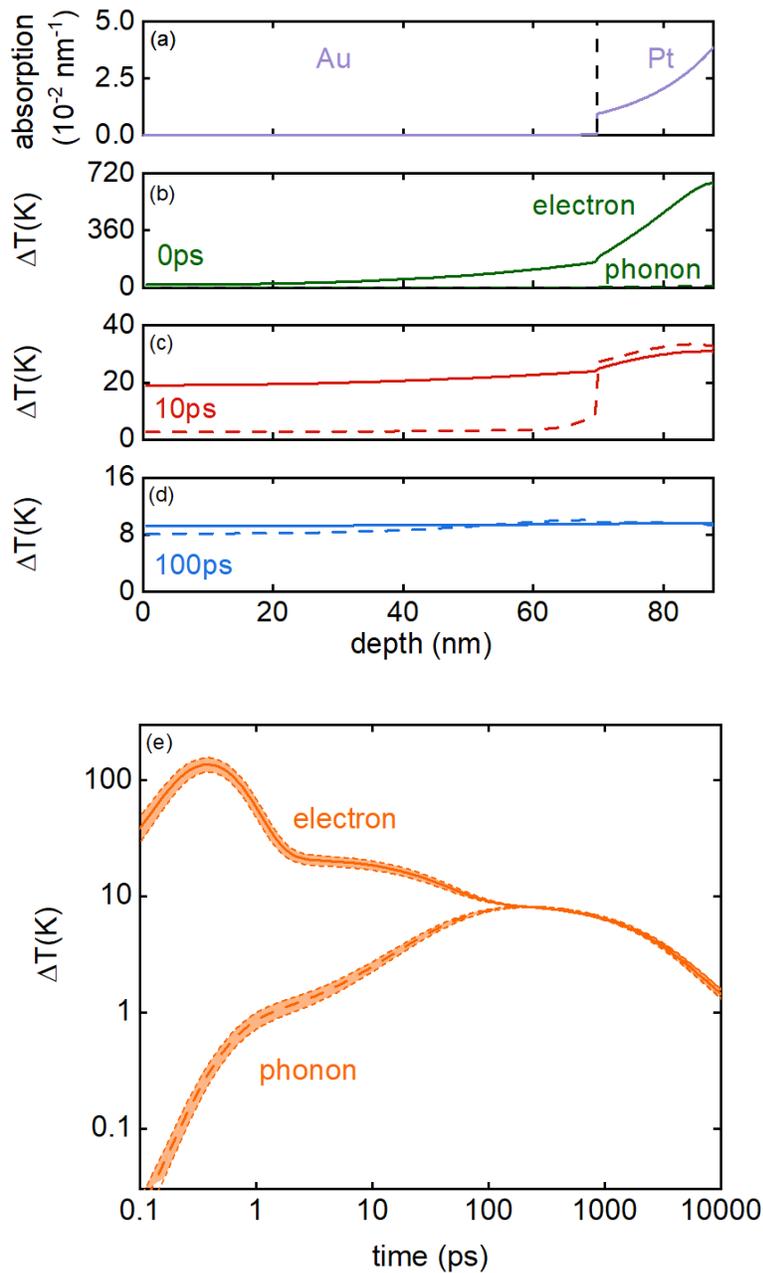

Figure 1. Two temperature model calculations for the Au/Pt bilayer for a pump wavelength of 783 nm. (a) Absorption of the pump beam vs. depth. (b-d) Electron temperature (solid lines) and phonon temperature (dashed lines) vs. depth at delay times of 0, 10, and 100 ps. Nonequilibrium between electrons and phonons in the Au layer persists for ~100 ps. (e) Electron and phonon temperature of the Au surface as a function of delay time. The shaded region represents the uncertainty in the electron and phonon temperature profiles due to uncertainties in thermal model parameters.

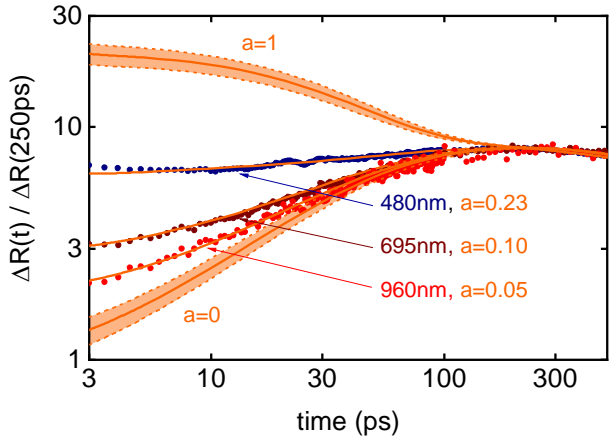

Figure 2. Pump/probe data for the Au reflectance at 480, 695, and 960 nm. The transient reflectance of the Au layer depends on wavelength due to changes in electron vs. phonon temperature sensitivity. Lines are thermal model predictions for the change in reflectance with different values for the electron temperature sensitivity parameter $a$. The shaded region represents our estimates for uncertainty in the electron and phonon temperature profiles due to uncertainties in thermal model parameters.

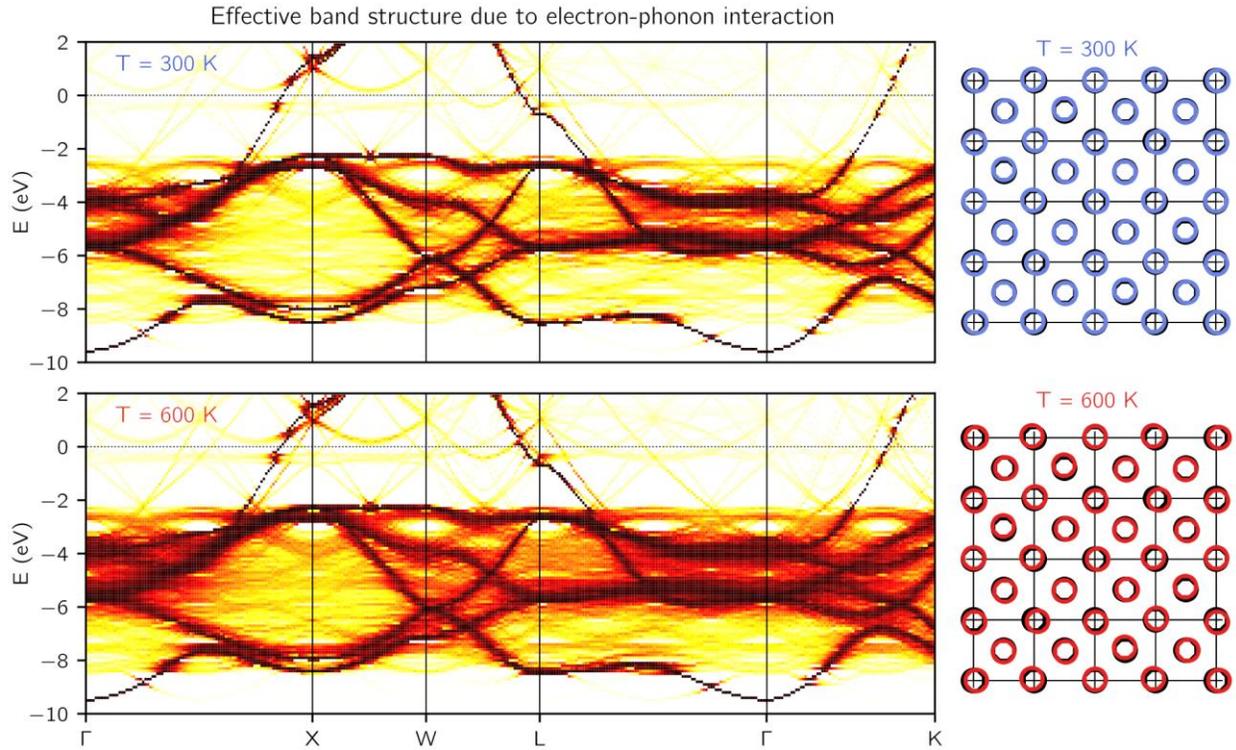

Figure 3. Band structure of gold, including electron-phonon interaction contribution, along high-symmetry path in the Brillouin zone for temperature of 300 K (top) and 600 K (bottom). Calculations are done in a 4x4x4 supercell with thermal displacements of atoms (blue and red circles in the right panels). The band structure is unfolded into a primitive unit cell, as indicated with color on the plot. Darker colors correspond to larger intensity.

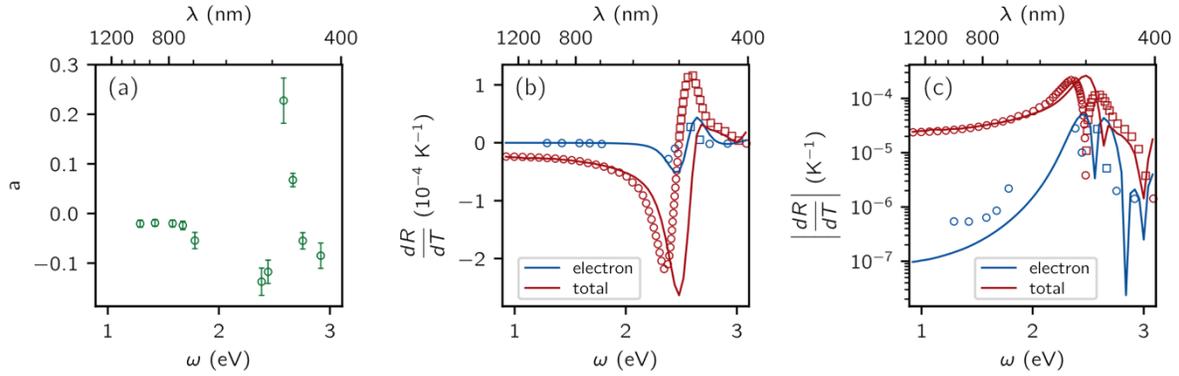

Figure 4. (a) Sensitivity of the thermoreflectance to the electron temperature as a function of photon energy, $a(\omega)$. (b-c) Electron (blue) and total (red) thermoreflectance as a function of photon energy. Measured data are shown with symbols while first-principles calculated results are shown with a lines.

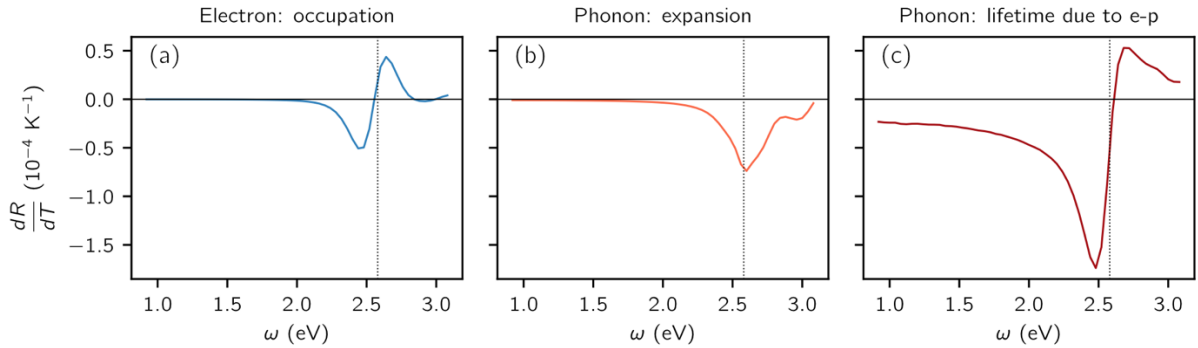

Figure 5. Decomposition of the calculated thermoreflectance as a function of photon energy. Three panels show (a) contributions from the electron occupation, (b) volume expansion, and (c) thermally displaced atoms. The primary effect of thermally displaced atoms is on the electron-phonon lifetime.

**Supplemental Material: Differentiating Contributions of Electrons and Phonons to the Thermoreflectance Spectra of Gold**

Kexin Liu[1], Xinping Shi[1], Frank Angeles[1], Ramya Mohan[2], Jon Gorchon[3], Sinisa Coh[1,2,a], Richard B. Wilson[1,2,b]

4) Mechanical Engineering, University of California, Riverside, CA, 92521, USA
5) Materials Science and Engineering, University of California, Riverside, CA, 92521, USA
6) Université de Lorraine, CNRS, IJL, F-54000 Nancy, France
c) sinisa.coh@gmail.com
d) rwilson@ucr.edu

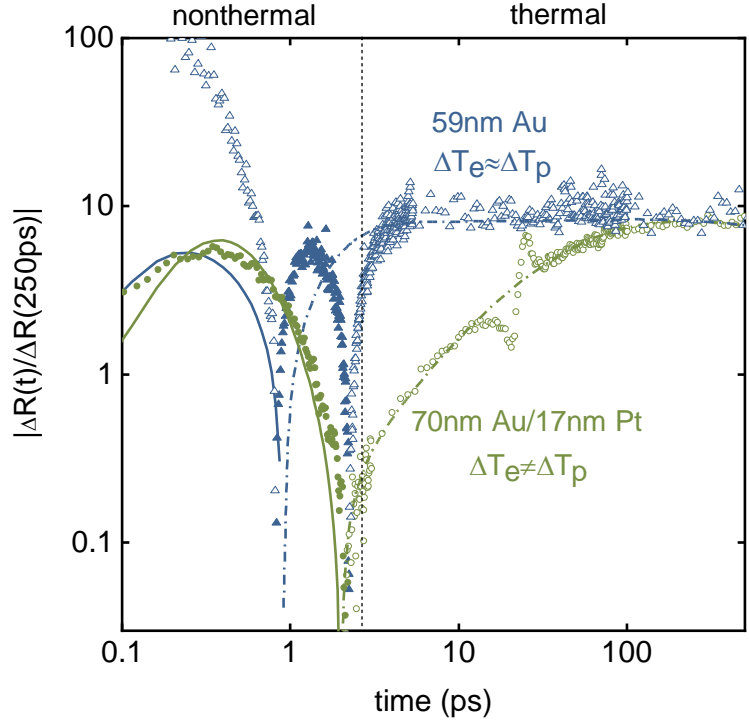

**Supplementary Figure 1.** Pump/probe data for Au and Au/Pt bilayer. Data is from a measurement with a pump wavelength of 900 nm, and a probe wavelength of 450 nm. Solid makers denote positive change in Au reflectance, while open markers are negative. The lines are the prediction of a two-temperature model as described in the main-text and below. We hypothesize that, at time scales less than 2 ps, the model predictions for the 59 nm Au film disagree with the data due to nonthermal effects and/or the effect of an acoustic strain wave. On timescales longer than 2 ps, electrons and phonons in the 60 nm Au layer are in equilibrium, and their contribution to the thermoreflectance signal cannot be differentiated. We overcome this obstacle by focusing our study on Au/Pt bilayers. In Au/Pt bilayers, the electron/phonon nonequilibrium persists for ~ 100 ps. The deviation between the model predictions and thermoreflectance data for the Pt/Au bilayer at ~30 ps is due to a longitudinal acoustic strain wave, which we use to measure the Pt and Au film thickness.

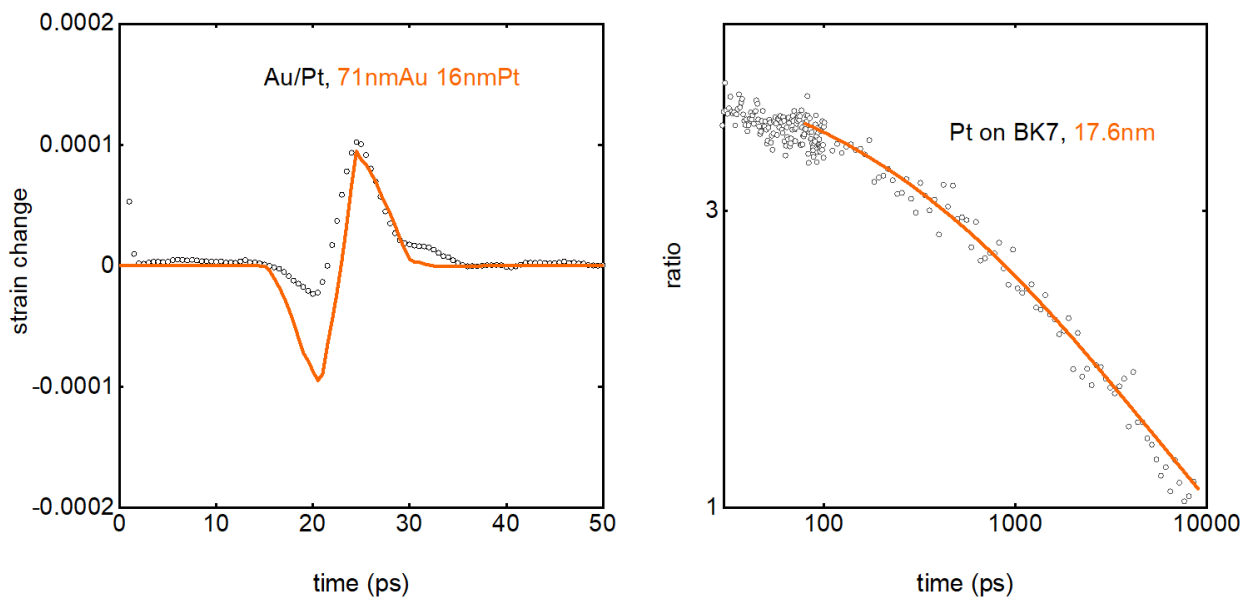

**Supplemental Figure 2.** (a) Picosecond acoustic data and acoustic model predictions. The picosecond acoustic data is arrived at by subtracting the thermal signal from our pump/probe data. We deduce the thickness of the layers by treating the Au and Pt thickness as fit parameters in the acoustic model. (b) TDTR measurement of a Pt film deposited on BK7 substrate with the same conditions as the Pt layer in the Au/Pt bilayer shown in (a). The line is a best fit to the data with a thermal model. We treated the Pt thickness as a fit parameter in the thermal model. We set the Pt film thickness for this bilayer to 16.8 nm, the average of the best-fit values in (a) and (b).

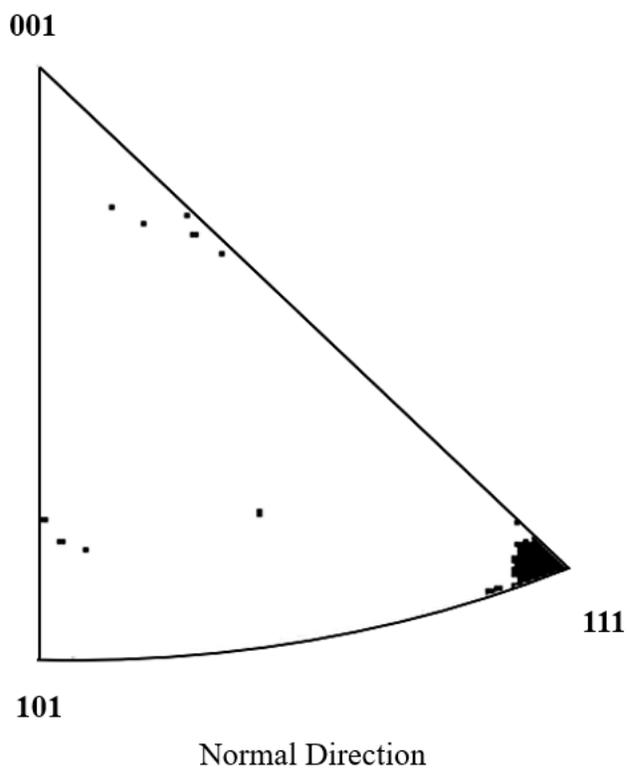

**Supplemental Figure 3.** A scatter plot of the EBSD Inverse Pole Figure of an Au/Pt/Sapphire film in the normal direction (thickness of Au layer ~70 nm). The data shows sharp crystalline texture in the [111] direction.

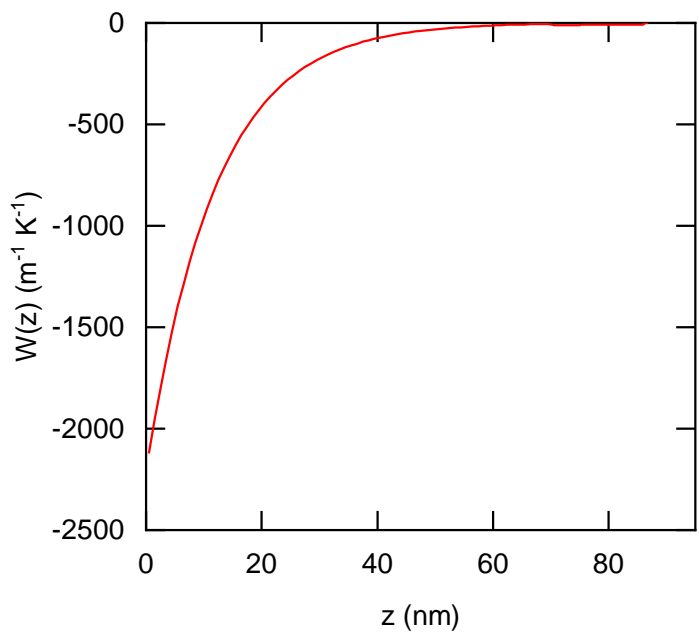

**Supplementary Figure 4.** Depth sensitivity of the thermoreflectance measurement of 70nm Au / 17nm Pt at a probe wavelength of 960 nm. We calculate the depth sensitivity using a multilayer reflectivity calculation [S1] and the thermo-optic coefficients for Pt and Au reported in Ref. [S2]. The area under the red line is the total thermoreflectance coefficient of the sample at a wavelength of 960 nm.

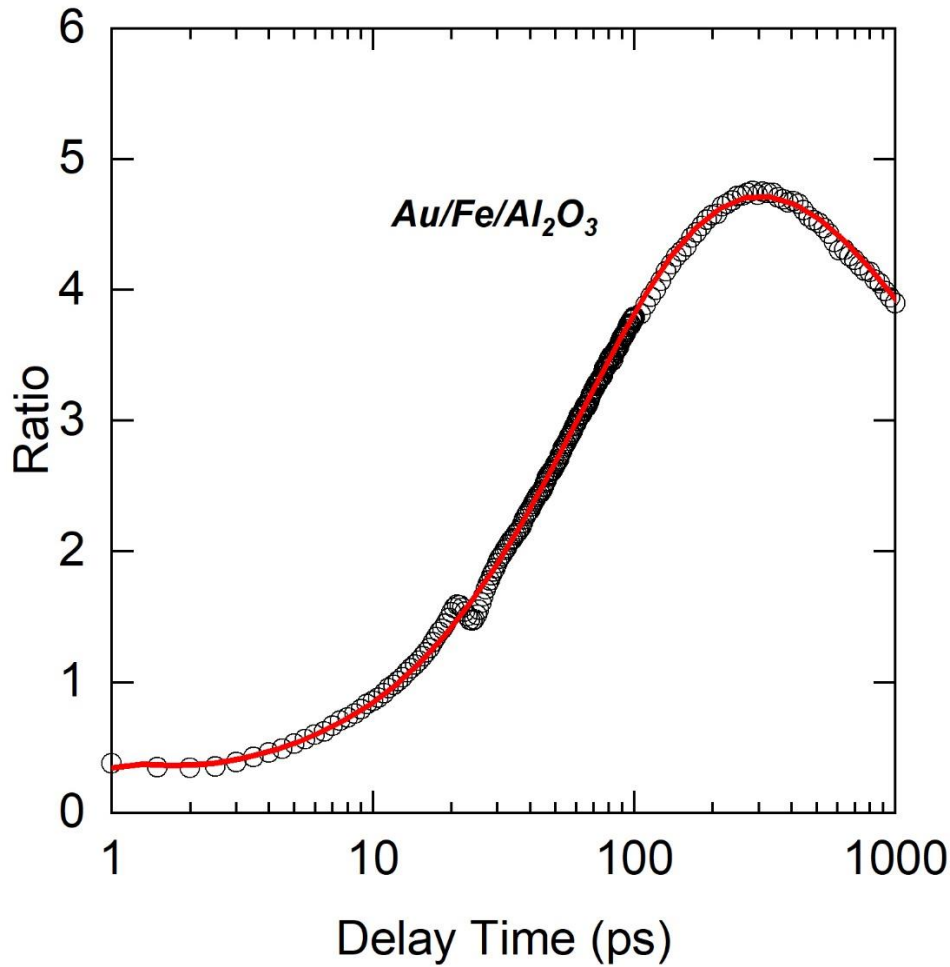

**Supplementary Figure 5.** Pump/probe data for a 73nm Au/ 18nm Fe bilayer sample. The data is proportional to change in reflectance of the Au at a wavelength of 783nm as function of pump probe time-delay. Lines are thermal model predictions for the change in reflectance with $|a| \approx 0.01$ and $|b| \approx 0.99$. The thermal model uses the following values for the thermal properties of Fe. An electron-phonon coupling parameter for Fe of $10^{18}$ W m$^{-3}$ K$^{-1}$. A phonon heat-capacity of 3.5 MJ m$^{-3}$ K$^{-1}$. An electron thermal conductivity (based on electrical resistivity measurements and WF-law) of 15 W m$^{-1}$ K$^{-1}$. A phonon thermal conductivity of 10 W m$^{-1}$ K$^{-1}$ (based on TDTR measurements of the total thermal conductivity for a separately prepared 200nm thick Fe film, and the WF-law prediction for total thermal conductivity.) An electronic heat-capacity of 0.02 MJ m$^{-3}$ K$^{-1}$.

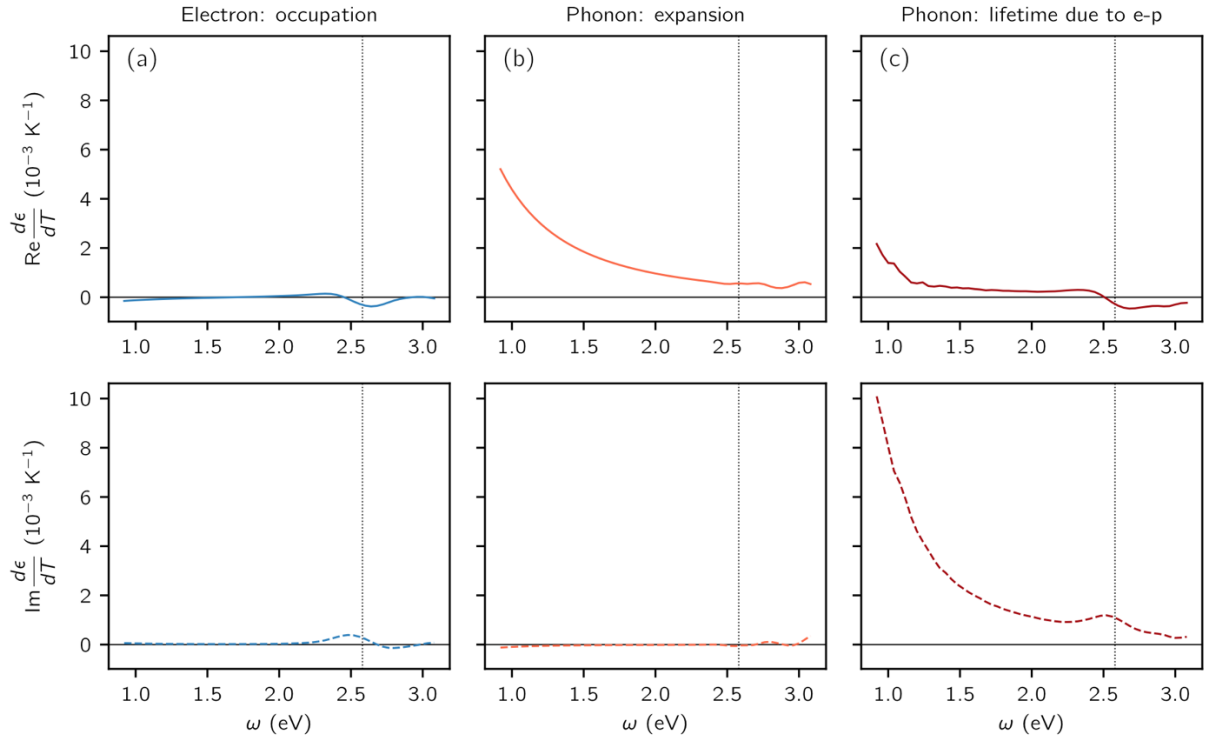

**Supplementary Figure 6.** Decomposition of calculated derivative of dielectric constant with respect to temperature as a function of photon energy. Top row shows the real part of the derivative, bottom row shows the imaginary part. Three columns correspond to contributions for the electron occupation, volume expansion, and thermal displacement of the atoms. The main effect of thermal displacement of the atoms is a change in the lifetime of carriers due to electron-phonon interactions.

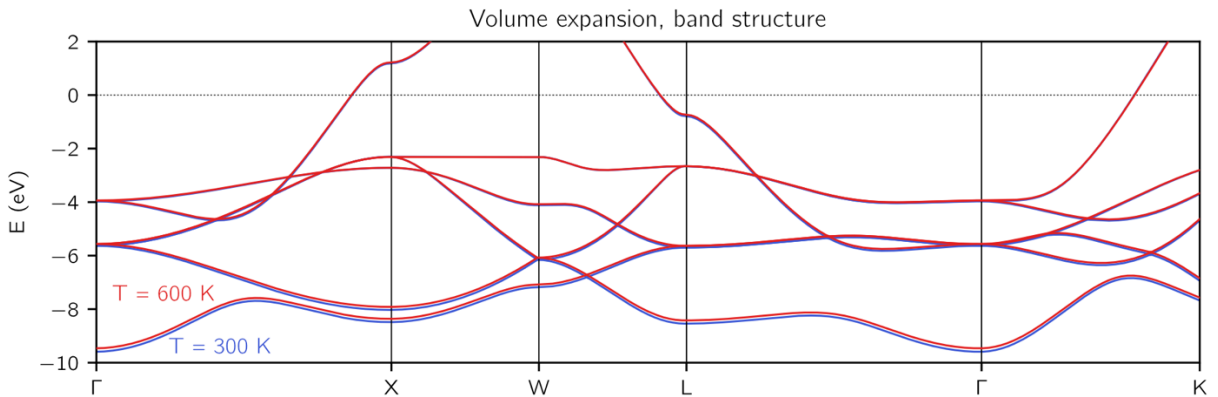

**Supplementary Figure 7.** Band structure of gold along high-symmetry path in the Brillouin zone for unit-cell volume at temperature of 300 K (blue) and 600 K (red).

**Sample Growth and Characterization**

The base pressure of the chamber was less than 4 x $10^{-7}$ torr prior to sputtering. During sputtering, the pressure was increased to 3.5 mTorr by introducing ultra-high purity Argon with a mass flow controller. We sputtered Pt and Au films from 1" and 2" inch targets at powers of 10 and 100 W, respectively.

Analysis of experimental data requires accurate measures of the Pt and Au film thicknesses. We use picosecond acoustics to measure the film thicknesses, see Supplementary Figure 2a. To further reduce uncertainty in the layer thickness, we also carefully calibrated the sputter deposition rates of the Au and Pt targets in the following manner. Before (after) deposition of the Au/Pt bilayer, we sputtered separate single-layer Au (Pt) films with conditions and times identical to those for the Au/Pt bilayers. We then used TDTR to determine the thicknesses of the Au and Pt single-layer samples by fitting the data with a thermal model whose only unknown parameter is film thickness, see Supplementary Figure 2b. These measurements of the "dummy" films gave thickness values within 10% of our estimates based on picosecond acoustic measurements of the bilayers themselves.

We measured the texture of the Au and Pt films with electron backscattering diffraction in a field-emission scanning electron microscope (Nova NanoSEM 450), equipped with an Oxford EBSD detector (NORDLYS Nano CCD Camera detector). We analyzed the EBSD maps and generated inverse pole figures (IPF) using the Oxford Instruments AztecSynergy software. Based on the scatter plots of the IPF in the normal direction, we find the film is highly textured in the [111] direction, see supplementary figure 3. We use the texture to estimate the speed of sound. The longitudinal speeds of sound in Pt and Au films in the [111] direction are 4280 and 3450 m/s, respectively.

**Depth Sensitivity of Thermoreflectance Signals**

Our experiments measure the temperature induced change of a sample's reflectance. In general, the change in reflectance of a sample will be a measure of the temperature-profile as a function of depth,

$$\Delta R(t) = \int_0^\infty \left[ W_e(z)\Delta T_e(z) + W_p(z)\Delta T_p(z) \right] dz . \qquad (S3)$$

Here, $W_e$ and $W_p$ are weight functions that determines the sensitivity of the thermoreflectance to the change in electron and phonon temperatures $\Delta T_e$ and $\Delta T_p$. When electrons and phonons are in equilibrium, Eq. (S3) simplifies to

$$\Delta R(t) = \int_0^\infty W(z) \Delta T(z) dz. \qquad (S4)$$

The weight functions decrease with increasing depth due to the finite optical penetration depth of light into metals.

To numerically calculate $W(z)$, we discretize the multilayer stack into 0.5 nm thick nodes. In other words, instead of performing a multilayer reflectivity calculation on a two layer system of air /70nm Au / 17nm Pt / sapphire, we perform the multilayer reflectivity calculation on a 177 layer system of air/[0.5nm Au]×140 / [0.5nm Pt]×34 / sapphire. The reflectivity of this system is defined as $R_0$. Then, to determine $W(z)$ at a specific depth $z0$, we modify the index of refraction of the node at depth $z_0$ from $n+ik$ to $\left(n+\frac{dn}{dT}(1K)\right)+i\left(k+\frac{dk}{dT}(1K)\right)$, and perform another multilayer reflectivity calcultaion. We define the reflectance of the modified system to be $R(z_0)$. Then, $W(z=z_0) = R_0 - R(z_0)$. We repeat this for all 177 layers to derive $W(z)$ at all depths.

Calculation of the electron and phonon weight functions for our samples, $W_e$ and $W_p$, requires knowledge of $\partial n/\partial T_e$ and $\partial k/dT_p$ for both Au and Pt, which we do not have. Alternatively, calculation of the weight function $W(z)$ requires only knowledge $dn/dT$ and $dk/dT$ for Au and Pt. We use the data reported in Ref. [S2] to calculate $W(z)$ for a (70nm Au)/(17nm Pt)/sapphire sample as a function of wavelength. We show an example plot for $W(z)$ at a wavelength of 960 nm in Supplemental Figure 4. Then, to analyze our data, we assume that $W_e(z) = a(\lambda)W(z)$ and $W_p(z) = b(\lambda)W(z)$. We then use Eq. (S3) and our two-temperature model predictions to calculate $\Delta R(t)$.

**Supplementary References**

S1.     Hecht, E., *Hecht optics.* Addison Wesley, 1998. **997**: p. 213-214.
S2.     Wilson, R., B.A. Apgar, L.W. Martin, and D.G. Cahill, *Thermoreflectance of metal transducers for optical pump-probe studies of thermal properties.* Optics express, 2012. **20**(27): p. 28829-28838.

# References

[1] R. Rosei, and D. W. Lynch, Thermomodulation Spectra of Al, Au, and Cu, Physical Review B **5**, 3883 (1972).
[2] R. Rosei, F. Antonangeli, and U. Grassano, d Bands position and width in gold from very low temperature thermomodulation measurements, Surface Science **37**, 689 (1973).
[3] R. Rosei, C. Culp, and J. Weaver, Temperature modulation of the optical transitions involving the Fermi surface in Ag: Experimental, Physical Review B **10**, 484 (1974).
[4] E. Colavita, S. Modesti, and R. Rosei, Thermoreflectance of Ag single crystals, Physical Review B **14**, 3415 (1976).
[5] P. Winsemius, F. Van Kampen, H. Lengkeek, and C. Van Went, Temperature dependence of the optical properties of Au, Ag and Cu, Journal of Physics F: Metal Physics **6**, 1583 (1976).
[6] P. B. Johnson, and R. W. Christy, Optical constants of copper and nickel as a function of temperature, Physical Review B **11**, 1315 (1975).
[7] J. H. Weaver, C. G. Olson, D. W. Lynch, and M. Piancentini, Thermoreflectance of Mo from 0.5 to 35 eV, Solid State Communications **16**, 163 (1975).
[8] W. Scouler, Temperature-modulated reflectance of gold from 2 to 10 eV, Physical Review Letters **18**, 445 (1967).
[9] N. E. Christensen, and B. O. Seraphin, Relativistic Band Calculation and the Optical Properties of Gold, Physical Review B **4**, 3321 (1971).
[10] C. A. Paddock, and G. L. Eesley, Transient thermoreflectance from thin metal films, Journal of Applied Physics **60**, 285 (1986).
[11] R. Wilson, B. A. Apgar, L. W. Martin, and D. G. Cahill, Thermoreflectance of metal transducers for optical pump-probe studies of thermal properties, Optics express **20**, 28829 (2012).
[12] H. Reddy, U. Guler, A. V. Kildishev, A. Boltasseva, and V. M. Shalaev, Temperature-dependent optical properties of gold thin films, Optical Materials Express **6**, 2776 (2016).
[13] P.-T. Shen, Y. Sivan, C.-W. Lin, H.-L. Liu, C.-W. Chang, and S.-W. Chu, Temperature- and roughness-dependent permittivity of annealed/unannealed gold films, Optics Express **24**, 19254 (2016).
[14] M. Xu, J.-Y. Yang, S. Zhang, and L. Liu, Role of electron-phonon coupling in finite-temperature dielectric functions of Au, Ag, and Cu, Physical Review B **96**, 115154 (2017).
[15] T. Favaloro, J.-H. Bahk, and A. Shakouri, Characterization of the temperature dependence of the thermoreflectance coefficient for conductive thin films, Review of Scientific Instruments **86**, 024903 (2015).
[16] J. Hohlfeld, S.-S. Wellershoff, J. Güdde, U. Conrad, V. Jähnke, and E. Matthias, Electron and lattice dynamics following optical excitation of metals, Chemical Physics **251**, 237 (2000).
[17] T. Heilpern, M. Manjare, A. O. Govorov, G. P. Wiederrecht, S. K. Gray, and H. Harutyunyan, Determination of hot carrier energy distributions from inversion of ultrafast pump-probe reflectivity measurements, Nature communications **9**, 1853 (2018).
[18] A. M. Brown, R. Sundararaman, P. Narang, A. M. Schwartzberg, W. A. Goddard, and H. A. Atwater, Experimental and Ab Initio Ultrafast Carrier Dynamics in Plasmonic Nanoparticles, Physical Review Letters **118**, 087401 (2017).
[19] M. Dresselhaus, Solid state physics part ii optical properties of solids, Lecture Notes (Massachusetts Institute of Technology, Cambridge, MA) **17** (2001).
[20] K. H. Bennemann, *Non-linear optics in metals* (Oxford University Press, 1998), 98.
[21] X. Wang, D. M. Riffe, Y.-S. Lee, and M. Downer, Time-resolved electron-temperature measurement in a highly excited gold target using femtosecond thermionic emission, Physical Review B **50**, 8016 (1994).

[22] S. Brorson, A. Kazeroonian, J. Moodera, D. Face, T. Cheng, E. Ippen, M. Dresselhaus, and G. Dresselhaus, Femtosecond room-temperature measurement of the electron-phonon coupling constant γ in metallic superconductors, Physical Review Letters **64**, 2172 (1990).

[23] P. Maldonado, K. Carva, M. Flammer, and P. M. Oppeneer, Theory of out-of-equilibrium ultrafast relaxation dynamics in metals, Physical Review B **96**, 174439 (2017).

[24] R. B. Wilson, and S. Coh, Parametric dependence of hot electron relaxation timescales on electron-electron and electron-phonon interaction strengths, Communications Physics **3**, 1 (2020).

[25] L. Waldecker, R. Bertoni, R. Ernstorfer, and J. Vorberger, Electron-Phonon Coupling and Energy Flow in a Simple Metal beyond the Two-Temperature Approximation, Physical Review X **6**, 021003 (2016).

[26] K. Sokolowski-Tinten, X. Shen, Q. Zheng, T. Chase, R. Coffee, M. Jerman, R. Li, M. Ligges, I. Makasyuk, and M. Mo, Electron-lattice energy relaxation in laser-excited thin-film Au-insulator heterostructures studied by ultrafast MeV electron diffraction, Structural Dynamics **4**, 054501 (2017).

[27] T. Chase, M. Trigo, A. Reid, R. Li, T. Vecchione, X. Shen, S. Weathersby, R. Coffee, N. Hartmann, and D. Reis, Ultrafast electron diffraction from non-equilibrium phonons in femtosecond laser heated Au films, Applied Physics Letters **108**, 041909 (2016).

[28] W. Wang, and D. G. Cahill, Limits to Thermal Transport in Nanoscale Metal Bilayers due to Weak Electron-Phonon Coupling in Au and Cu, Physical Review Letters **109**, 175503 (2012).

[29] G.-M. Choi, R. B. Wilson, and D. G. Cahill, Indirect heating of Pt by short-pulse laser irradiation of Au in a nanoscale Pt/Au bilayer, Physical Review B **89**, 064307 (2014).

[30] H. Jang, J. Kimling, and D. G. Cahill, Nonequilibrium heat transport in Pt and Ru probed by an ultrathin Co thermometer, Physical Review B **101**, 064304 (2020).

[31] See Supplemental Material at [URL] for additional pump/probe data, details on sample characterization, and thermoreflectance modelling.

[32] M. J. Gomez, K. Liu, J. G. Lee, and R. B. Wilson, High sensitivity pump–probe measurements of magnetic, thermal, and acoustic phenomena with a spectrally tunable oscillator, Review of Scientific Instruments **91**, 023905 (2020).

[33] R. Wilson, and D. G. Cahill, Experimental validation of the interfacial form of the Wiedemann-Franz law, Physical Review Letters **108**, 255901 (2012).

[34] H. Kurt, R. Loloee, K. Eid, W. Pratt Jr, and J. Bass, Spin-memory loss at 4.2 K in sputtered Pd and Pt and at Pd/Cu and Pt/Cu interfaces, Applied Physics Letters **81**, 4787 (2002).

[35] C. Galinon, K. Tewolde, R. Loloee, W.-C. Chiang, S. Olson, H. Kurt, W. Pratt Jr, J. Bass, P. Xu, and K. Xia, Pd/Ag and Pd/Au interface specific resistances and interfacial spin flipping, Applied Physics Letters **86**, 182502 (2005).

[36] R. Wilson, B. A. Apgar, W.-P. Hsieh, L. W. Martin, and D. G. Cahill, Thermal conductance of strongly bonded metal-oxide interfaces, Physical Review B **91**, 115414 (2015).

[37] C. Monachon, L. Weber, and C. Dames, Thermal boundary conductance: A materials science perspective, Annual Review of Materials Research **46**, 433 (2016).

[38] E. Hecht, Hecht optics, Addison Wesley **997**, 213 (1998).

[39] S. L. Adler, Quantum Theory of the Dielectric Constant in Real Solids, Physical Review **126**, 413 (1962).

[40] N. Wiser, Dielectric Constant with Local Field Effects Included, Physical Review **129**, 62 (1963).

[41] M. Zacharias, and F. Giustino, One-shot calculation of temperature-dependent optical spectra and phonon-induced band-gap renormalization, Physical Review B **94**, 075125 (2016).

[42] M. Zacharias, and F. Giustino, Theory of the special displacement method for electronic structure calculations at finite temperature, Physical Review Research **2**, 013357 (2020).

[43] A. M. Brown, R. Sundararaman, P. Narang, W. A. Goddard, and H. A. Atwater, Ab initio phonon coupling and optical response of hot electrons in plasmonic metals, Physical Review B **94**, 075120 (2016).

[44] J. Sólyom, *Fundamentals of the Physics of Solids: Volume 3-Normal, Broken-Symmetry, and Correlated Systems* (Springer Science & Business Media, 2010), Vol. 3.

[45] I. Campillo, J. M. Pitarke, A. Rubio, and P. M. Echenique, Role of occupied $d$ states in the relaxation of hot electrons in Au, Physical Review B **62**, 1500 (2000).

[46] F. Ladstädter, U. Hohenester, P. Puschnig, and C. Ambrosch-Draxl, First-principles calculation of hot-electron scattering in metals, Physical Review B **70**, 235125 (2004).

[47] M. Bernardi, J. Mustafa, J. B. Neaton, and S. G. Louie, Theory and computation of hot carriers generated by surface plasmon polaritons in noble metals, Nature communications **6** (2015).

[48] R. Bauer, A. Schmid, P. Pavone, and D. Strauch, Electron-phonon coupling in the metallic elements Al, Au, Na, and Nb: A first-principles study, Physical Review B **57**, 11276 (1998).

[49] J. Cao, Y. Gao, H. E. Elsayed-Ali, R. J. D. Miller, and D. A. Mantell, Femtosecond photoemission study of ultrafast electron dynamics in single-crystal Au(111) films, Physical Review B **58**, 10948 (1998).

[50] M. Aeschlimann, M. Bauer, S. Pawlik, R. Knorren, G. Bouzerar, and K. Bennemann, Transport and dynamics of optically excited electrons in metals, Applied Physics A **71**, 485 (2000).

[51] P. Giannozzi, O. Baseggio, P. Bonfà, D. Brunato, R. Car, I. Carnimeo, C. Cavazzoni, S. De Gironcoli, P. Delugas, and F. Ferrari Ruffino, Quantum ESPRESSO toward the exascale, The Journal of Chemical Physics **152**, 154105 (2020).

[52] M. Otter, Temperaturabhängigkeit der optischen konstanten massiver metalle, Zeitschrift für Physik **161**, 539 (1961).

[53] J. Pudell, A. Maznev, M. Herzog, M. Kronseder, C. Back, G. Malinowski, A. Von Reppert, and M. Bargheer, Layer specific observation of slow thermal equilibration in ultrathin metallic nanostructures by femtosecond X-ray diffraction, Nature communications **9**, 3335 (2018).

[54] O. Lozan, R. Sundararaman, B. Ea-Kim, J.-M. Rampnoux, P. Narang, S. Dilhaire, and P. Lalanne, Increased rise time of electron temperature during adiabatic plasmon focusing, Nature Communications **8**, 1656 (2017).

[55] M.-N. Su, C. J. Ciccarino, S. Kumar, P. D. Dongare, S. A. Hosseini Jebeli, D. Renard, Y. Zhang, B. Ostovar, W.-S. Chang, P. Nordlander, N. J. Halas, R. Sundararaman, P. Narang, and S. Link, Ultrafast Electron Dynamics in Single Aluminum Nanostructures, Nano Letters **19**, 3091 (2019).

[56] G. Tagliabue, A. S. Jermyn, R. Sundararaman, A. J. Welch, J. S. DuChene, R. Pala, A. R. Davoyan, P. Narang, and H. A. Atwater, Quantifying the role of surface plasmon excitation and hot carrier transport in plasmonic devices, Nature Communications **9**, 3394 (2018).

[57] E. L. Radue, J. A. Tomko, A. Giri, J. L. Braun, X. Zhou, O. V. Prezhdo, E. L. Runnerstrom, J.-P. Maria, and P. E. Hopkins, Hot Electron Thermoreflectance Coefficient of Gold during Electron–Phonon Nonequilibrium, ACS Photonics **5**, 4880 (2018).

[58] A. Block, M. Liebel, R. Yu, M. Spector, Y. Sivan, F. J. García de Abajo, and N. F. van Hulst, Tracking ultrafast hot-electron diffusion in space and time by ultrafast thermomodulation microscopy, Science Advances **5**, eaav8965 (2019).

[59] R. B. Wilson, J. Gorchon, Y. Yang, C.-H. Lambert, S. Salahuddin, and J. Bokor, Ultrafast magnetic switching of GdFeCo with electronic heat currents, Physical Review B **95**, 180409 (2017).

[60] G.-M. Choi, B.-C. Min, K.-J. Lee, and D. G. Cahill, Spin current generated by thermally driven ultrafast demagnetization, Nature communications **5**, 4334 (2014).

[61] G.-M. Choi, C.-H. Moon, B.-C. Min, K.-J. Lee, and D. G. Cahill, Thermal spin-transfer torque driven by the spin-dependent Seebeck effect in metallic spin-valves, Nature Physics **11**, 576 (2015).

[62] J. Kimling, G.-M. Choi, J. T. Brangham, T. Matalla-Wagner, T. Huebner, T. Kuschel, F. Yang, and D. G. Cahill, Picosecond spin Seebeck effect, Physical Review Letters **118**, 057201 (2017).

[63] T. S. Seifert, S. Jaiswal, J. Barker, S. T. Weber, I. Razdolski, J. Cramer, O. Gueckstock, S. F. Maehrlein, L. Nadvornik, S. Watanabe, C. Ciccarelli, A. Melnikov, G. Jakob, M. Münzenberg, S. T. B. Goennenwein, G. Woltersdorf, B. Rethfeld, P. W. Brouwer, M. Wolf, M. Kläui, and T. Kampfrath, Femtosecond formation

dynamics of the spin Seebeck effect revealed by terahertz spectroscopy, Nature communications **9**, 2899 (2018).

[64] K. T. Regner, L. C. Wei, and J. A. Malen, Interpretation of thermoreflectance measurements with a two-temperature model including non-surface heat deposition, Journal of Applied Physics **118**, 235101 (2015).

[65] R. Wilson, J. P. Feser, G. T. Hohensee, and D. G. Cahill, Two-channel model for nonequilibrium thermal transport in pump-probe experiments, Physical Review B **88**, 144305 (2013).

[66] D. G. Cahill, P. V. Braun, G. Chen, D. R. Clarke, S. Fan, K. E. Goodson, P. Keblinski, W. P. King, G. D. Mahan, and A. Majumdar, Nanoscale thermal transport. II. 2003–2012, Applied Physics Reviews **1**, 011305 (2014).

[67] J. Kimling, and D. G. Cahill, Spin diffusion induced by pulsed-laser heating and the role of spin heat accumulation, Physical Review B **95**, 014402 (2017).

[68] J. Kimling, R. Wilson, K. Rott, J. Kimling, G. Reiss, and D. G. Cahill, Spin-dependent thermal transport perpendicular to the planes of Co/Cu multilayers, Physical Review B **91**, 144405 (2015).

[69] J. Gorchon, R. B. Wilson, Y. Yang, A. Pattabi, J. Y. Chen, L. He, J. P. Wang, M. Li, and J. Bokor, Role of electron and phonon temperatures in the helicity-independent all-optical switching of GdFeCo, Physical Review B **94**, 184406 (2016).

[70] A. A. Maznev, J. A. Johnson, and K. A. Nelson, Non-equilibrium transient thermal grating relaxation in metal, Journal of Applied Physics **109**, 073517 (2011).

[71] J. P. Freedman, R. F. Davis, and J. A. Malen, Nondiffusive electron transport in metals: A two-temperature Boltzmann transport equation analysis of thermoreflectance experiments, Physical Review B **99**, 054308 (2019).

[72] Y. Sivan, and M. Spector, Ultrafast dynamics of optically-induced heat gratings in metals--more complicated than expected, arXiv preprint arXiv:1909.03122 (2019).